\documentstyle[11pt,paspconf]{article}
\begin{document}

\title{Accretion Disk Line Emission in AGN: A Devil's Advocacy}

\author{Jack W. Sulentic}
\affil{Department of Physics and Astronomy, University of Alabama,
Tuscaloosa, AL, USA 35487}

\author{Paolo Marziani}
\affil{Osservatorio Astronomico, vicolo dell'Osservatorio 5,
Padova, Italy}

\author{Deborah Dultzin-Hacyan}
\affil{Instituto de Astronomia, UNAM, Apdo. 70-264,
Mexico D. F. 04510, Mexico}

\begin{abstract}
We review the evidence for AGN optical and X-ray broad line emission
from an accretion disk. We argue that there is little, if any,
statistical evidence to support this assertion. The inconsistency is
strongest for the rare class of Balmer profiles that show double
peaks.  The line profiles predicted by a simple illuminated disk model
are often incompatible with the observations.  We suggest that the Fe
K$\alpha$\ line in Seyfert 1 galaxies, where a broad line is most often
and most strongly detected, is actually a composite of two lines both
with Gaussian profiles; one narrow/unshifted and the other
broad/redshifted.   \end{abstract}

\section{Introduction}

One can ask two questions about accretion disks in connection with
active galactic nuclei (AGN). Do AGN host a supermassive black hole
(BH) and associated accretion disk (AD)? Can we see line emission
directly from the putative AD? The latter question has come to
represent the best hope for obtaining a positive answer to the former
one (a ``proof'' of the BH/AD paradigm) since the central broad line
region (BLR) of AGN will remain unresolved even with space telescopes
for the foreseeable future (for NGC~1068 at 12 Mpc, 1pc = 0.02
arcsec).  Excellent theoretical arguments have been made both to show
the necessity for a BH/AD and also to explain some serious
observational problems (see, for example, Collin-Souffrin et al.
1990):
(i) a medium for producing the optical FeII emission, (ii) a means of
explaining the implied density and observed profile velocity
differences between high (HIL: e.g., CIV $\lambda$ 1549) and low (LIL:
e.g., Balmer lines) ionization broad lines.

The hope of detecting accretion disk line emission received a boost
when the double peaked H$\alpha$ profile for Arp 102B was shown to be
well fit by a relativistic Keplerian disk model (Chen \& Halpern 1989;
Chen, Halpern \& Filippenko 1989). New encouragement came later from
the detection of broad and redshifted Fe K$\alpha$\ line profiles with
the ASCA satellite (e.g., Tanaka et al. 1995).  Seyfert 1 galaxies, in
particular, appear to show a broad and complex line near 6.4 keV which
has been fit with a wide range of disk models (see, for example, Nandra
et al.  1997a,b).  In this case we are presumably seeing line emission
from the innermost regions of the disk where relativistic effects are
more pronounced. In \S\ 2 we consider four categories of problems for
optical emission from accretion disks. This is  followed in \S\ 3 by a
discussion of problems with the interpretation of the X-ray line
emission.  In \S\ 4 we describe the problems arising if both Fe
K$\alpha$\ and Balmer lines are assumed to arise in a disk.

\section{Double-Peaked Optical Lines: from AD?}

\subsection{Double-Peaked Profiles are Rare}
\nobreak

Disk model fits to the Arp 102B Balmer lines were remarkably good and
accounted for both the redshifted base of the profile and the brighter
blue peak. A subsequent attempt to find more sources with profiles like
Arp 102B however was disappointing. Only 12/94 sources studied in a
(biased) sample of radio loud sources with peculiar profiles (like Arp
102B) could be satisfactorily fit with a disk model (Eracleous \&\
Halpern 1994). This was especially disturbing because the disk
inclination derived from the fit to Arp 102B implied an intermediate
viewing angle where many sources should be found. More generally,
double peaked profiles are found only among the radio-loud AGN which
represent no more than $\sim$15\% of all AGN (Kellermann et al.
1989).  This implies that double peaked profiles are found in less than
1\% of AGN which is consistent with the results of spectroscopic
surveys.  The rarity of sources similar to Arp 102B imply that  double
peaked sources are: (1) pathological, (2) miraculous or (3)
orientationally constrained. There is no indication that the
double-peaked sources differ significantly from other radio-loud AGN
except in the shape of their line profiles (Eracleous \&\ Halpern
1994).  Explanation (2) requires either that most sources, where double
peaked lines should be seen, have an additional source of line emission
that fills in the gap between the peaks, or that most disks emit lines
at much larger radii in the disk.  The ``additional source''
possibility cannot provide a general solution because double-peaked
profiles, with or without a filled in center, are broader than the
majority of AGN.  In either case then the detection of a few sources
line Arp 102B becomes disturbingly miraculous. We prefer an explanation
where double peak rarity is explained because they represent sources
seen at near pole-on orientation. In that case the peaks might arise
from opposite sides of a biconical outflow (Zheng et al. 1990, Sulentic
et al.  1995a). While bicone models are not without difficulties (see,
for example, Livio \&\ Xu 1997), they provide the simplest explanation
for the rarity of double-peaked profiles.

\subsection{Double-peaked Profiles are not Strong Fe Emitters}
\nobreak

Double-peaked sources appear to be exclusively radio-loud AGN while
optical FeII emission is found to be, on average, stronger in the
radio-quiet majority of AGN (Marziani et al. 1996). Examination of the
H$\beta$ region in the spectra of sources like Arp 102B suggest a lower
than (radio-loud) average Fe II optical strength. We estimate  FeII
optical/H$\beta\le 0.2$ for Arp 102B while we found a mean value near
0.8 for radio-loud sources in general (Marziani et al. 1996).
Ironically, one of the original reasons for advocating disk emission
was because it provided a suitable medium for FeII line production.
There is also evidence that FeII optical and Balmer emission correlate
in the sense that both H$\beta$ and the FeII lines broaden in the same
way -- suggesting that they might arise from the same emitting region
(Phillips 1978; Boroson \& Green 1992). If one argues that Balmer lines
are produced in an accretion disk it would therefore be difficult not
to expect significant FeII emission, especially, from double-peaked
sources where it is argued that the Balmer lines arise in the disk.

\subsection{The Peaks Vary Out of Phase}
\nobreak

The red and blue peaks in the double-peaked sources vary out of phase
(3C390.3: Gaskell 1988, Zheng, Veilleux \&\ Grandi 1991, Arp 102B:
Miller \& Peterson 1990)). Pictor A (Sulentic et al.  1995a) originally
showed single-peak Balmer profiles, but both red and blue peaks
appeared independently over the past ten years.  The double peaks
appear to be transient in NGC 1097 (Storchi-Bergmann et al. 1997).
Various, more complex, disk models (spotted, warped, elliptical
disk/ring) have been advanced to account for the peak variability. We
find the ring model (Storchi-Bergmann et al. 1997) particularly
attractive, but not in support of disk emission.  It would replace the
disk with an emitting accretion ring that might remove a major problem
for the bicone scenario (Livio \&\ Xu 1997) involving the difficulty
for observing emission from the far-side clouds.

\subsection{Statistical Confrontations Between AD Models and Spectra}
\nobreak

\begin{figure}[tbp]
\vspace{15truecm}
\includegraphics{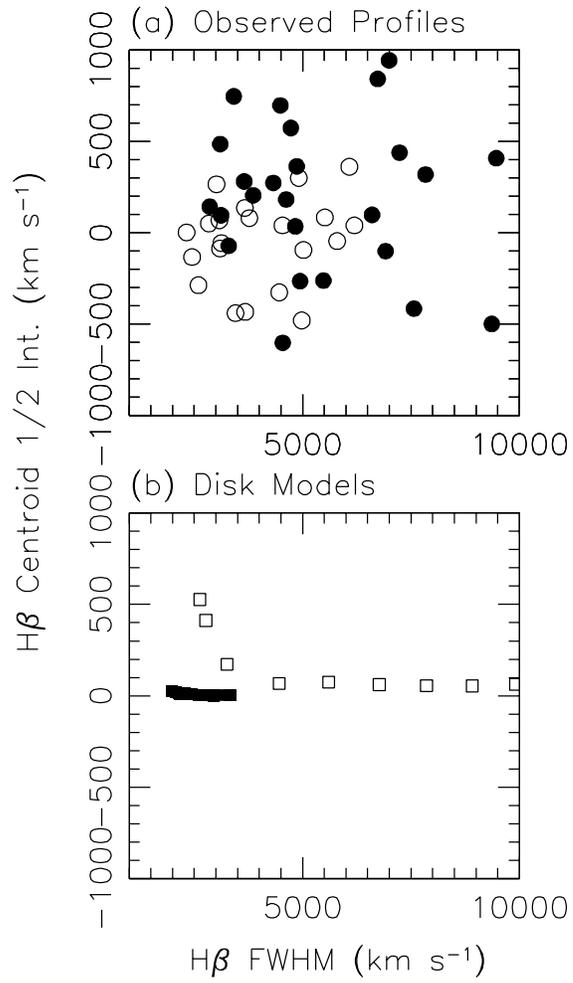}
\caption{ Distribution of Balmer line centroid shift vs. FWHM for: (a)
radio-quiet (open circles) and radio-loud (filled circles) AGN and (b)
disk model predictions.
Filled boxes  and open boxes are for model 1 and  model 3 by Sulentic
et al.
1998, respectively.}
\label{jwsfig1}
\end{figure}

The first studies of line profile shifts and asymmetries (Gaskell 1982;
Wilkes
1984) found evidence for a systematic blueshift of the HIL relative to
the LIL.
This difference was cited as an argument for a two component BLR where
the LIL,
with mean shift near zero (assumed to be the AGN rest frame) arose in a
disk
(Collin-Souffrin et al. 1980, 1982).  Larger studies utilizing HST
observations
of HIL in the same sources where the LIL were measured (Sulentic et al.
1995b;
Marziani et al. 1996) reveal that the HIL blueshift is present only in
radio-quiet sources where double-peaked profiles have not been
observed. The
statistical results of such studies are generally bad for disk models.
We
earlier (Sulentic et al.  1990) made a general comparison between model
predictions and observed Balmer line shift/asymmetry properties. The
general
trends predicted by disk models (blue shifted/asymmetric peaks and
redshifted
line profile bases are not observed. Figure~\ref{jwsfig1} shows the
half
maximum  line shift vs. width parameter space for both radio-loud and
radio-quiet
sources (Marziani et al. 1996) compared to the predictions of the disk
model
summarized in \S\ 4.

The situation for disk emission is therefore bad in both observational
and disk model parameter spaces. In the latter case, the scatter in
disk model fits (inclination, inner emitting radius and emissivity
power-law) show no convergence towards a single physical model (see,
for example, Sulentic et al. 1998).  The models that have the greatest
potential to produce BLR emission in the bulk of (especially
radio-quiet) quasars must move the emitting region of the disk to
$\sim$10$^4$ r$_g$ ($r_g$ = GM/c$^2$, the gravitational radius of the
central black hole).  This has the unsatisfactory result of making the
double-peak sources miraculous in a new way -- the only sources where
emission is primarily observed from the inner parts of the disk (i.e.,
a
special emissivity law).

\section{X-ray Lines from AD?}

The ASCA mission has opened up a new spectroscopic window for the study
of AGN. Emission at 6.4 keV was detected previously but without
sufficient spectral resolution or S/N to  characterize the line
profile. MCG -06-30-15 has become almost a prototype for the
characteristic line profile of the 6.4 keV feature in Seyfert 1
galaxies. The discussion now shifts from radio-loud to radio-quiet
sources.  The picture for X-ray line emission in the optical
double-peakers is not yet clear. 3C390.3 has been detected (Eracleous
et al. 1996) while Pictor A shows no detectable Fe line emission
(Halpern et al. 1997).

Fe line emission at 6.4 keV is not really double peaked. It almost
always shows
a strong blue (high energy) peak near 6.4 keV and a very broad red wing
extending down to 4-5 keV. This is generally consistent with disk
models where
the X-ray line arises in the innermost (R$\le$ 20 r$_g$) part of the
AD. Line
emission can arise at three different energies in this part of the
spectrum:
(i) at 6.4keV, from fluorescence reflection by ``cold'' Fe; (ii) in the
range
6.45-6.7 keV,  from highly ionized Fe  or (iii) at 6.97 keV,  from
H-like Fe.
The general consensus that cold Fe dominates the emission is supported
by the
frequent detection of an expected reflection component in the
continuum. In the
following subsections we consider the problems with a disk
interpretation for
the 6.4 keV line.

\subsection{A Two Component Profile For the 6.4 keV Line?}
\nobreak

Nandra et al. (1997a,b) recently reported a study of 16 of the (mostly
Seyfert 1) sources with a strong 6.4 keV line. They generated an
average profile from this data both with and without MCG -06-30-15 and
NGC 4151. Both average profiles show a 6.4 keV narrow component and a
broader redshifted feature. The redshifted feature is better defined in
the average profiles and looks Gaussian in shape. In the presence of a
complex line one can: (1) look for an obvious deconvolution into
symmetrical components before (2) try to fit it with a more complex
model based upon a popular but unproven hypothesis. This approach
suggests that the 6.4 keV line may be a composite of two independent
(roughly symmetric and Gaussian) features: (1) a narrow (FWHM = 0.24
keV)
unshifted (E$_n$ = 6.4$\pm$0.05 keV) line and (2) a broad (FWHM = 1.6
keV)
redshifted (E$_n$ = 5.9$\pm$0.1 keV) line. We recently showed
(Sulentic,
Marziani \&\ Calvani 1998) that the $\chi^2$ of our own model fit to
the low stage
was no better than that obtained from a Gaussian fit with E$_n$=5.2 keV
and $\sigma$= 0.75 keV. If the Gaussian interpretation is
correct then it immediately invalidates all published disk model fits.

An important justification for adopting this deconvolution involves the
blue side of the Fe profile. Most disk models explored so far show a
sharp decrease in intensity on that side of the line due to Doppler
boosting.  As the data has improved, the observations have shown a
softening of this wing. The excess blue wing flux is well seen in
MCG -05-23-16 (Weaver et al. 1997) and, most significantly, in the
average spectra of Nandra et al.  (1997a,b). While we cannot rule out a
contribution from ``hot'' Fe in this high energy wing, it appears to
form a smooth extension, under the 6.4 keV narrow peak, of the broad
redshifted component of the line. We have recently discussed possible
origins
(and references) for the Fe emission viewed as two independent lines
(Sulentic
et al.  1998; Sulentic, Marziani  \&\ Calvani 1998). No obvious disk
model can
account for the Gaussian components interpreted as independent lines
unless the
X-ray emission arises at the outer edge of the disk (10$^4$--10$^5$
r$_g$).

\subsection{Variability in MCG -06-30-15: More Trouble for AD}
\nobreak

The extensive observations of the Seyfert 1 source MCG -06-30-15 reveal
considerable variability in the 6.4 keV line. High, medium and low
variability line profiles  have been derived (Iwasawa et al.  1996b).
The medium stage of variability resembles the typical Fe profile
characterized by the average Fe spectrum (Nandra et al. 1997a,b).
During
the low phase the narrow 6.4 keV peak appears to disappear completely
while at high phase the broad red component either remains constant or
decreases in equivalent width. This is consistent with the above
suggestion that the broad and narrow components of the line are
independent. The situation may be more complex, especially if one
removes the X-ray data associated with the most extreme variability,
but there are scarcely enough photons to allow a clear interpretation
beyond what is summarized here.

We attempted (Sulentic, Marziani \&\ Calvani 1998) to find an
illuminated AD model
(details in final section) that could reproduce all three
stages of the Fe profile variability in MCG -06-30-15. We obtained
physically
consistent fits by varying the index $\xi$ of the power law emissivity
between 0.7 (high phase) and 3.0 (low phase). Actually the fit to the
high
phase profile was too high. This reflects the fact that we could only
obtain
self consistent fits to all phases by requiring that the broad
redshifted component remains constant. Iwasawa et al. (1996b) present
evidence for almost an order of magnitude change in the broad component
EW  between high and low phase. In any case the fit requires a nearly
maximally rotating BH which is not otherwise physically justified.

\subsection{Why is the Peak always at 6.4 keV?}
\nobreak
The position of the Doppler boosted peak in disk models will change
with the inclination of the AD to our line of sight. In a unification
model we therefore expect to see the peak move to higher energy as the
disk is viewed at higher inclinations. The data for ``AD-like''
profiles currently shows a strong concentration of narrow blue peaks at
6.4 keV. In the Nandra et al. (1997a,b) sample no source shows a peak
higher than 6.45 keV. Admittedly the statistics are still poor, but the
wavelength (energy) ``stability'' of the blue peak is more consistent
with the interpretation that it is an independent line than with the
predictions of disk models.

\subsection{Recent Results for Seyfert 2 Sources}
\nobreak

 In a ``unification'' picture, the Seyfert 2 galaxies are Seyfert 1's
with near edge-on inclination where the BLR is hidden by an obscuring
torus. Such a scenario can be coupled with the assumption that the Fe
line arises from a disk. If the obscuring torus is optically thick at
6.4 keV then we would expect to detect little or no line emission. With
the exception of IRAS 1832-5926 (Iwasawa et al. 1996a)  and MCG
-05-23-16
(Weaver et al. 1997), initial results suggested that Seyfert 2 emission
was confined to a narrow feature at 6.4 keV. This already presented a
challenge for disk models because the AD fits to Sy1 profiles left
little residual flux in the peak for attribution to a second source
that would be common to both Seyfert classes (analogous to the narrow
emission lines). The most likely source of the non-AD Seyfert component
would have been the obscuring torus which might contribute significant
cold Fe.  The variability results for MCG -06-30-15 however suggest
that the
blue peak responds almost instantaneously to continuum changes. This is
inconsistent with Fe emission from a torus believed to lie up to a
parsec from the central engine.

\begin{figure}[tbp]
\vspace{8truecm}
\includegraphics{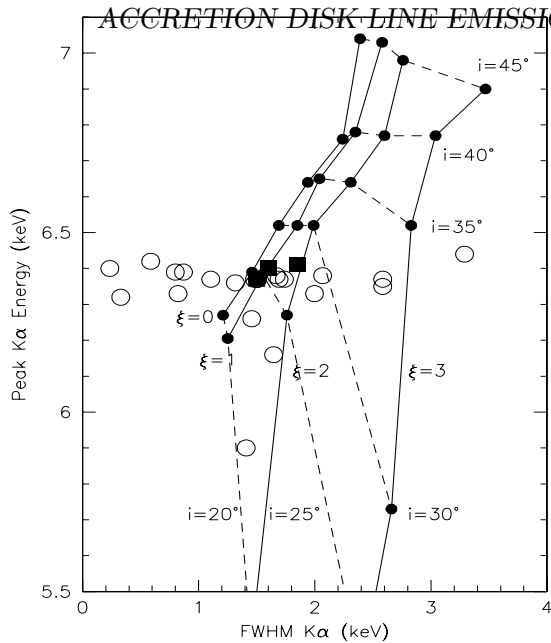}
\caption{Comparison of disk model predictions for Fe K$\alpha$\ line
narrow
peak centroid and broad profile FWHM and observed values for Seyfert
galaxies.
Filled squares are data points for Seyfert 2 galaxies, open circles for
Seyfert
1.}
\label{jwsfig2}
\end{figure}

A further challenge to disk models has emerged from a study of the Fe
line profiles for 25 Sy 2 galaxies (Turner et al. 1998). The average Sy
2 profile appears to show the same shape (although sometimes much less
flux) as the average Sy 1 which is at variance with unification
predictions. Two of us have argued that many or most Seyfert 2 galaxies
are not edge-on Seyfert 1's on independent grounds (see Dultzin-Hacyan
et al. 1998).  This result amplifies the problem for AD that was raised
in the previous section. It is independent of the role of absorption in
reconciling the observed profile similarities. The narrow peak centroid
is the most robust measure of the line profiles and it again shows a
strong clustering at 6.4 keV.  Figure~\ref{jwsfig2}  compares  model
predictions (solid circles, for different values of inclination and and
of power-law emissivity index $\xi$) with the observed Fe narrow peak
energy and full profile FWHM for Seyfert 1 (open circles; Nandra et
al.
1997a,b) sources and  Seyfert 2 group averages (see Turner et al.
1998).

The concentration of narrow peak centroids at 6.4 keV represent a
strong challenge to an AD origin for the line emission.

\section{An Attempt at ``Unification'' of Optical and X-ray Line
Emission}

The parallel arguments for AD line emission from the Balmer and X-ray
Fe lines suggest that it would be useful to explore illumination models
that might simultaneously produce both lines. One of the biggest
criticisms that can be leveled against model fits to optical and X-ray
line profiles is the lack of convergence of the solutions in the
parameter space. In most cases each line profile is fit without
constraint on any of the principal disk parameters. If these fail to
provide an adequate solution then various ``epicyclic'' modifications
of the disk are employed. The disk model parameter space is effectively
three dimensional since, disk inclination, inner emitting radius and
index of the emissivity law govern the profile shape. We recently
reported an attempt at developing a simple disk illumination model that
would produce both Balmer lines and the 6.4 keV feature (Sulentic et
al.  1998).

\subsection{The Illuminated Disk Model}
\nobreak

We tried to make the fewest possible assumptions in developing this
model so we started with a standard geometrically thin $\alpha$ disk
with corrective terms given by Novikov \& Thorne (1973) for a Kerr BH
with angular momentum per unit mass a= 0.998. This is probably
reasonable for a Seyfert 1 source like  MCG -06-30-15 where the
Eddington ratio is L$_{bol}$/L$_{Edd}$ $\leq$0.34. A BH mass
M$\geq$10$^7$M$_{\odot}$ and $\dot{\rm M}\approx$ 2$\times$10$^{23}$g
s$^{-1}$ was adopted. We assumed that the disk is illuminated by
radiation produced at the center and scattered both by a sphere with
radius 100 r$_g$ and by a spherical halo of free electrons with radius
$\approx$10$^6$r$_g$, both with electron-scattering optical depth
$\tau_{es}=0.25$. We chose two extreme models for the continuum shape
that should represent over- and underestimates of the real continuum.
The flux scale was provided by observations of MCG -06-30-15. The
radial disk
structure appropriate for  a Kerr metric was employed with
a = 0.998 GM/c (Thorne 1974).  The CLOUDY photoionization code (Ferland
1996) was used. The cold Fe K$\alpha$\ line profile was computed
following
Fanton et al. (1997) which incorporates all relevant relativistic
effects for a Kerr metric.

\subsection{Results of the Disk Model}
\nobreak
We derived line emissivity  as a function of disk radius for both Fe
K$\alpha$\ and H$\alpha$\ using CLOUDY. The bulk of the cold Fe
emission originated between 1.25 and 20 r$_g$, with the total
contribution of hot to cold Fe $\le$0.3, while H$\alpha$ peaks at
10$^3$--10$^4$ r$_g$. These predictions are generally consistent with
many previous model fits to the Fe line. They also suggest that AGN
should strongly favor symmetric single-peaked Balmer line profiles. The
double peaked profiles are again found to have a ``miraculous''
character even if the broad line emission from most sources is
dominated by disk emission. They are miraculous in the sense that they
would represent a very rare population of AGN with H$\alpha$ emission
originating from radii an order of magnitude closer to the center than
expected (e.g., 350 -- 1000r$_g$ for Arp 102B; Eracleous \&\ Halpern
1994).  The latter limit on the outer emitting radius is also
completely {\it ad hoc} since there is no physical reason to truncate
the optical line emission so close to the center.

\begin{figure}[tbp]
\vspace{8truecm}
\includegraphics{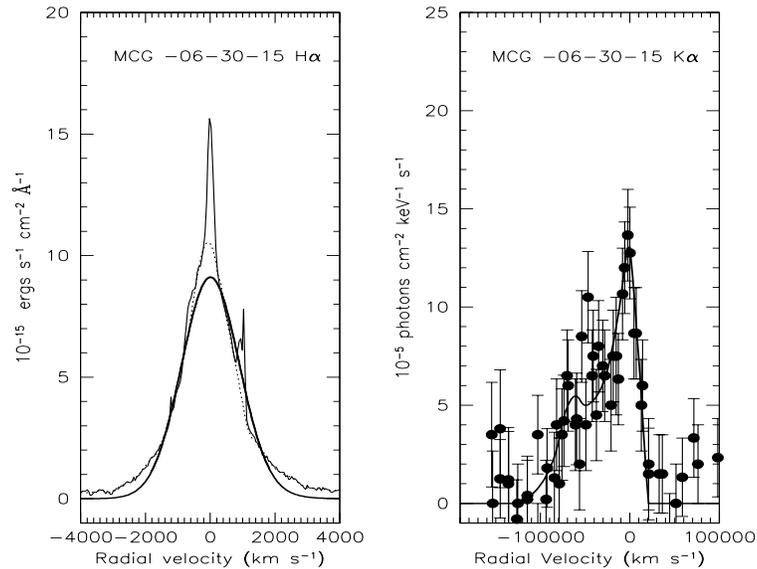}
\caption{Optical H$\alpha$\ (right) and X-ray Fe K$\alpha$\ (left) line
profiles of MCG -06-30-15 with best fit disk models superposed. The
disk model
profile for K$\alpha$\ has been computed with an improved version of
the code
by Fanton et al. (1997) and is slightly different from that shown in
Sulentic,
Marziani, \&\ Calvani (1998).
\label{jwsfig3}}
\end{figure}

We obtained  optical H$\alpha$\ spectra for 15 of the sources that show
broad Fe  K$\alpha$\ emission (Sulentic et al. 1998).  This enables us
to compare our model predictions  about the line profiles of both Fe
K$\alpha$\ and H$\alpha$.  In addition we can use the line profiles to
derive the disk inclination which is the single free parameter in the
models. We discussed earlier the model profiles for MCG -06-30-15 in
three stages of variability.  Predicted profiles for Fe are generally
in better agreement than those for H$\alpha$. This may be more the
result of the poor S/N of the Fe spectra than anything else. The
agreement between optical profiles and model predictions is poor. The
fit is particularly bad in the wings where the disk emission might be
expected to dominate. In addition the observed Balmer line shifts and
asymmetries are not predicted by the model (see Calvani et al. 1997 for
a possible solution).  Another problem involves the inconsistencies
between disk inclinations derived from optical and X-ray spectra for
the same source. MCG -06-30-15  shows a very narrow H$\alpha$\ profile
that constrains it to a small inclination angle while Fe K$\alpha$\ is
constrained to a value near 30$^{\circ}$. Figure~\ref{jwsfig3} shows
the observed optical and X-ray line profiles with our best model fits
superposed.

Few sources show agreement in the
derived inclinations but, of course, many X-ray derived inclinations
are poorly constrained.

We conclude that, aside from some individual fits to a few unusual
sources, there is no compelling observation evidence for line emission
from the putative accretion disk in AGN.

\end{document}